\documentstyle[12pt]{article}
\newcommand{\slp}{\raise.15ex\hbox{$/$}\kern-.57em\hbox{$\partial$}}
\newcommand{\sla}{\raise.15ex\hbox{$/$}\kern-.57em\hbox{$a$}}
\newcommand{\slA}{\raise.15ex\hbox{$/$}\kern-.57em\hbox{$A$}}
\newcommand{\slB}{\raise.15ex\hbox{$/$}\kern-.57em\hbox{$B$}}
\newcommand{\slb}{\raise.15ex\hbox{$/$}\kern-.57em\hbox{$b$}}
\newcommand{\slW}{\raise.15ex\hbox{$/$}\kern-.57em\hbox{$W$}}
\newcommand{\be}{\begin{equation}}
\newcommand{\ee}{\end{equation}}
\newcommand{\bear}{\begin{eqnarray}}
\newcommand{\ear}{\end{eqnarray}}

\newcommand{\eins}{  1\!{\rm l}  }

\begin{document}
\begin{flushright}
HD--THEP--95--6
\end{flushright}
\quad\\
\vspace{1.8cm}
\begin{center}
{\bf\LARGE Fermionic Coset Realization of Primaries}\\
\medskip
{\bf\LARGE in Critical Statistical Models}\\
\vspace{1cm}
D. C. Cabra\footnote[1]{On leave of absence from the Universidad
Nacional de La Plata, Argentina} and K. D. Rothe\\
\bigskip
Institut  f\"ur Theoretische Physik\\
Universit\"at Heidelberg\\
Philosophenweg 16, D-69120 Heidelberg
\end{center}
\vspace{2.0cm}
\begin{abstract}
We obtain a fermionic coset realization of the primaries of minimal
unitary
models and show how their four-point functions may be calculated
by the use of a reduction formula. We illustrate the construction
for the Ising model, where we obtain an explicit realization of
the energy operator, Onsager fermions, as well as of the order
and disorder operators realizing the dual algebra, in terms of
constrained Dirac fermions. The four-point correlators of these
operators are shown to agree with those obtained by other methods.
\end{abstract}
\newpage
\section{Introduction}

The study of two-dimensional Conformal Quantum Field Theory (CQFT)
has received much attention in the last ten years, following the
pioneering
work of Belavin, Polyakov, and Zamolodchikov (BPZ) \cite{BPZ}.
Interest
in this subject has been promoted by the relation of 2D-CQFT's to
two-dimensional statistical systems near second-order
phase transitions as well as in their relevance in the study of
classical
solutions of string theories\footnote{See ref. \cite{Ibb} for an extensive
coverage and list of references.}.

Based on the general approach developed in
\cite{BPZ}, Friedan, Qiu and
Shenker (FQS) have shown \cite{FQS}
that unitarity of the representation of the
Virasoro algebra (VA) restricts the possible values of the
central charge
$c$,  and the conformal dimensions $h$ of the fields to the range
$c\geq 0, h\geq0$, with $c$ and $h$ taking only the discrete values
\be\label{1.1}
c=1-\frac{6}{(k+2)(k+3)},\quad k=1,2...\ee
\be\label{1.1b}
h_{p,q}^{(k)}=\frac{(p(k+3)-q(k+2))^2-1}{4(k+2)(k+3)}\ee
if $0<c<1$. These values of $c$ characterize the so-called ``minimal
unitary models''. In particular, FQS have shown
that for the first four values
of $c$ in the series (\ref{1.1}),
$c=\frac{1}{2},\frac{7}{10},\frac{4}{5},\frac{6}{7}$,
the conformal dimensions given by (\ref{1.1b}) coincide with the
critical exponents of the Ising Model (IM),
Tricritical-IM, 3-state Potts
Model and Tricritcal Potts Model, respectively.
About the same time
it has further been shown \cite{H} that
the critical points of RSOS models
\cite{ABF} provide in fact particular realizations
of all members in the
discrete series.

The first attempt to obtain explicit Quantum Field theoretic
realizations
of these models was made in ref. \cite{GKO}, where the coset
construction
was used to obtain new representations of the Virasoro algebra (VA).
In particular
it was shown that the coset $G/H$, with $G=SU(2)_k\times SU(2)_1$ and
$H=SU(2)_{k+1}$, provide realizations of the
Virasoro algebra of minimal unitary models. The possibility of an
explicit
realization of these coset models by gauging WZW Lagrangians was also
suggested.

Since then, the search for Lagrangian formulations of the coset
construction
have received much attention  \cite{GK}, \cite{KS},
\cite{BRS}, \cite{CMR}.

In refs. \cite{GK}, \cite{KS} it has been shown that the central
charges of suitably gauged WZW models coincide with the central
charges of the GKO coset models. The same conclusion has been
reached from the study of gauged fermionic models
\cite{BRS}, \cite{CMR}.
In \cite{KS} it was further shown that the energy-momentum tensor
of the gauged WZW theory coincides with the one
in the GKO construction
in the physical (BRST-invariant) subspace.

As far as the minimal unitary models are concerned,
some specific primary
fields were constructed in the bosonic formulation
by Gawedzki and Kupiainen (GK) \cite{GK} and a general
method for obtaining the other primaries has been given.
In the formulation
of \cite{GK} the 4-point functions of these primaries can be evaluated
using a decoupled picture. A study of the full operator content of
both bosonic and fermionic gauged theories,
including the relation between
their primary fields and the corresponding scaling
fields in statistical models
is, however, lacking.

In the present paper we partially fill this gap
and show how to realize
some of the primary fields, in the unitary minimal models
in the framework of
the fermionic coset construction \cite{BRS}, \cite{CMR}.
By representing
the partition function of the fermionic coset formulation
as a product of
``decoupled'' sectors, we show that the primary fields are given by
BRST-invariant composites of the decoupled fields.
We furthermore show
how to obtain the correlators of the primaries in this picture.

In the particular case of $k=1,c=\frac{1}{2}$
(corresponding to the IM)
our fermionic construction goes
beyond the bosonic one of GK as we give a
field-theoretic realization
of all operators (energy-, order-, disorder-operators
and Majorana
fermions) of the critical Ising model, as well as
of the corresponding
order-disorder algebra, in terms of BRST-invariant local composites of
decoupled fields involving massless bosons and Dirac fermions.
We further show that all
(mixed and unmixed) 4-point correlation functions agree
with the expected
results.

The paper is organized as follows:

In section 2 we present the partition
function for the GKO coset $(SU(2)_k\times SU(2)_1)/SU(2)_{k+1}$ in the
fermionic coset formulation, and show that it can be written as the
product of the partition functions of free Dirac fermions,
negative level WZW fields and ghosts.

The computation of the four-point function of
primaries in the ``decoupled picture'' requires the evaluation
of four-point correlators of negative level
$SU(n)$-WZW fields in arbitrary
representations.
For positive level $SU(n)$-WZW fields in the fundamental
representation the general solution
has been given by Knizhnik and Zamolodchikov (KZ) \cite{KZ}. The
computation of higher-spin 4-point correlators has, however,
been restricted to
$SU(2)$ WZW fields \cite{FZ}. Making use of the
equivalence between the
$SU(n)_N$ WZW theory and the fermionic coset $U(nN)/(SU(N)_n\times
U(1))$ \cite{NS}, we derive in section 3 a
``reduction formula'' allowing one
to calculate the correlators of primaries in terms of WZW correlators
belonging to $SU(2)$.

In section 4 we then specialize to the case of $k=1$, which is expected
to be equivalent to the critical two-dimensional Ising model \cite{IZ}.
We begin by identifying the primary fields corresponding to the energy
and spin operators and show that their respective 4-point functions
agree with known results \cite{KC}, \cite{LP}, \cite{BPZ}.
We next identify the disorder operator by requiring
the usual order-disorder
algebra and show that the 4-point function involving
both order and disorder
operators again agrees with known results \cite{BPZ}.
A brief summary of these results has been reported recently \cite{CR}.
The computations
of the correlators
are significantly simplified by the use of the reduction formula derived
in section 3. We conclude this section by identifying the
Onsager \cite{On} (Majorana) fermions.

In section 5 we discuss how the above construction of the primaries may be
generalized to the whole FQS series, and conclude in section 6.

\section{Fermionic Coset Realization of the Minimal Unitary Series}
\setcounter{equation}{0}

As was proposed in ref. \cite{P}, the fermionic
realization of the $G/H$ coset
model can be obtained from a free fermion Lagrangian with
symmetry $G$ $(G=
U(N),O(N),..)$ by freezing the degrees of freedom associated with the
subgroup $H$,  imposing the conditions
\be\label{2.1}
J_\mu^{(H)}|phys>=0\ee
where $J_\mu^{(H)}$ are the conserved currents
associated with the group
$H$. In the path integral formalism (\ref{2.1})
is implemented by gauging
the $H$-symmetry of the fermionic Lagrangian.

As mentioned in the introduction, the minimal unitary series can be
represented by the coset models
\be\label{2.2}
G/H=\frac{SU(2)_k\times SU(2)_1}{SU(2)_{k+1}}\ee
One obtains a fermionic representation of
a model with symmetry group
$SU(2)_k\times SU(2)_1$ in terms of cosets,
by making use of the general
equivalence \cite{NS}
\be\label{2.3}
SU(N)_k\hat=\frac{U(Nk)}{SU(k)_N\times U(1)}\ee
where the l.h.s. stands for the realization of a
$SU(N)\ WZW$-model of
level $k$, and the r.h.s. is realized by a theory of
$N\times k$ free Dirac
fermions $\psi^{i\alpha}(i=1,...N$ and $\alpha=1,...,k)$, with the
$SU(k)$ currents $\psi^{\dagger i\alpha}\gamma_\mu
T^a_{\alpha\beta}\psi^{i\beta},
\ (T^a,a=1,...,k^2-1$ the $SU(k)$ generators) and the
$U(1)$ currents $\psi^{\dagger i\alpha}\gamma_\mu\psi^{i\alpha}$ freezed
by gauging the respective symmetry groups. In this way one obtains a
coset representation of the numerator of (\ref{2.2}), and one is led to
make the identification
\be\label{2.4}
\frac{SU(2)_k\times SU(2)_1}{SU(2)_{k+1}}=
\frac{\frac{U(2k)}{SU(k)_2\times U(1)}
\times\frac{U(2)}{U(1)}}{SU(2)_{k+1}}\ee
where the group $SU(2)_{k+1}$ is moded out by
further freezing the currents
$\psi^{\dagger i\alpha}\gamma_\mu t^a_{ij}\psi^{j\alpha}+
\chi^{\dagger i\alpha}\gamma_\mu t^a_{ij}\chi^j$, (with
$t^a$ the $SU(2)$ generators),
which satisfy
a Kac-Moody algebra of level $k+1$.
The identification (2.4)
receives support from the equality of central charges
as well as from the current
correlation functions in the corresponding Lagrangian
formulation \cite{GK}, \cite{CMR}.

According to the above prescriptions, the Lagrangian
realizing the r.h.s.
of (\ref{2.4}) is given by
\bear\label{2.5}
{\cal L}^{(k)}&=&\frac{1}{\sqrt2\pi}\psi^{\dagger i\alpha}
\left((\slp+i\sla)
\delta_{ij}\delta_{\alpha\beta}+i\slB^a
T^a_{\alpha\beta}\delta_{ij}
+i\slA^at^a_{ij}\delta_{\alpha\beta}\right)
\psi^{j\beta}+\nonumber\\
&+&\frac{1}{\sqrt2\pi}\chi^{\dagger i\alpha}\left((\slp+i\slb)
\delta_{ij}+i\slA^at^a_{ij}
\right)\chi^j\ear
where $i,j=1,2;\ \alpha,\beta=1,2,...,k$, and the gauge fields act
as Lagrange multipliers implementing the conditions (2.1).

In order to arrive at a ``decoupled'' description, we change variables by
writing
\bear\label{2.6}
a\ =&i(\bar\partial h_a)h_a^{-1},\qquad \bar a\ =&i(\partial
\bar h_a)\bar h_a^{-1}\nonumber\\
b\ =&i(\bar\partial h_b)h_b^{-1},\qquad \bar b\ =&i(\partial
\bar h_b)\bar h_b^{-1}\nonumber\\
A\ =&i(\bar\partial g_A)g_A^{-1},\qquad \bar A\ =&i(\partial
\bar g_A)\bar g_A^{-1}\nonumber\\
B\ =&i(\bar\partial g_B)g_B^{-1},\qquad \bar B\ =&i(\partial
\bar g_B)\bar g_B^{-1}\nonumber\\
&&\nonumber\\
\psi_1\ =&h_ag_Ag_B\psi^{(0)}_1,\qquad
\psi^\dagger_2\ =&\psi^{(0)\dagger}_2
(h_ag_Ag_B)^{-1}\nonumber\\
\psi_2\ =&\bar h_a\bar g_A\bar g_B\psi^{(0)}_2,
\qquad\psi^\dagger_1\ =&\psi^{(0)\dagger}_1
(\bar h_a\bar g_A\bar g_B)^{-1}\nonumber\\
&&\nonumber\\
\chi_1\ =&h_bg_A\chi_1^{(0)},\qquad\chi_2^\dagger\ =
&\chi_2^{(0)^\dagger}
(h_bg_A)^{-1}
\nonumber\\
\chi_2\ =&\bar h_b\bar g_A\chi_2^{(0)},\qquad\chi_1^\dagger\ =&
\chi_1^{(0)^\dagger}
(\bar h_b\bar g_A)^{-1}\ear
where $A_\mu=A^a_\mu t^a$ and $B_\mu=B^a_\mu T^a$.
Under the gauge transformations $W_\mu\to GW_\mu G^{-1}+G\partial
_\mu G^{-1}$, with $W_\mu$ standing for $A_\mu,B_\mu,a_\mu$ and $b_\mu$,
the products
\be\label{2.7}
\tilde g_I=g^{-1}_I\bar g_I,\ I=A,B;\quad\tilde h_i=h_i^{-1}\bar h_i,
\quad i=a,b\ee
remain invariant. Parametrizing $\tilde h_i$ by
\be\label{2.8} \tilde h_i=e^{-2\phi_i}\ee
taking due account of the Jacobians of the respective transformations
\cite{PW}, \cite{Fi}
(see also \cite{CMR}),  and factoring out the infinite gauge volume
$\Omega=\int {\cal D}g_A{\cal D}g_B\newline
{\cal D}h_a{\cal D} h_b$
one arrives at the following decoupled form for the partition
function associated with the Lagrangian (\ref{2.5}):
\be\label{2.9}
Z_{\frac{SU(2)_k\times SU(2)_1}{SU(2)_{k+1}}}=Z_FZ_BZ_{WZW}Z_{gh}\ee
where
\newpage
\bear\label{2.10}
 Z_F&=&\int{\cal D}\psi^{(0)}{\cal D}\psi^{(0)\dagger}
{\cal D}\chi^{(0)}{\cal D}\chi^{(0)\dagger}\exp
\left(-\frac{1}{\pi}\int(\psi^{(0)
\dagger}_2\bar\partial\psi_1+\psi^{(0)\dagger}_1
\partial\psi^{(0)}_2)\right)\cdot\nonumber\\
&&\qquad\cdot\exp\left(-\frac{1}{\pi}\int(\chi^{(0)\dagger}
_2\bar\partial\chi_1^{(0)}+\chi^{(0)\dagger}_1
\partial\chi^{(0)}_2)\right)\\
Z_B&=&\int{\cal D}\phi_a{\cal D}\phi_b
\exp\left(\frac{k}{\pi}\int \phi_a\Delta
\phi_a\right)\exp\left(\frac{1}{\pi}\int\phi_b\Delta\phi_b\right)
\nonumber\\
Z_{WZW}&=&\int{\cal D}\tilde g_A{\cal D}\tilde g_B
\exp((k+5)W[\tilde g_A])
\exp((2k+2)W[\tilde g_B])\nonumber\ear
and where $Z_{gh}$ is the partition functions of $SU(k),
\ SU(2),\ U(1)$
and $U(1)$ decoupled ghost fields. The
explicit form of the ghost partition function will not be required
at this time. $W[g]$ is the $WZW$ functional \cite{Wi}
\bear\label{2.11}
W[g]&=&\frac{1}{16\pi}\int d^2 x tr(\partial_\mu g
\partial^\mu g^{-1})+\nonumber\\
&&+\frac{1}{24\pi}\int d^3 y\varepsilon_{ijk} tr(g^{-1}
\partial_i g g^{-1}\partial_j gg^{-1}\partial_k g).\ear
where the second integral is over the three-dimensional
ball with space time as boundary.
The total central charge is obtained by adding the
individual contributions. The central charge associated with a
$WZW$ field of level $K$
is given by the well-known formula \cite{KZ}
\be\label{2.12}
c=\frac{ K\dim G}{K+C_V}\ee
where $C_V$ is the Casimir of the corresponding symmetry group $G$ in
the fundamental representation: $f^{abc}f^{a'bc}=C_V\delta^{aa'}$.
Thus we have for the individual contributions in $Z_{WZW}$,
\be\label{2.13}
c^{(A)}_{WZW}=\frac{3(k+5)}{k+3},\ c^{(B)}_{WZW}=
\frac{2(k+1)(k^2-1)}{k+2}.\ee
Adding to these central charges the contributions
$c_F=2k+2,\ c_B=2$ and $c_{gh}=-2k^2-8$, we obtain (1.1),
thus giving support to the identification (\ref{2.4}).

Note that the $WZW$-sectors have negative levels $-2(k+1)$ and $-(k+5)$,
respectively, which taken by themselves
would imply the presence of negative norm states. Unitarity is, however,
restored by taking into account the other sectors.
Although the different sectors appear decoupled on the level of the
partition function (2.9), they are in fact coupled via the BRST
quantization conditions, the observables of the theory being required
to be BRST invariant \cite{KS}.
In terms of the variables of the gauged Lagrangian (\ref{2.5}),
this amounts to considering only gauge invariant composites of
these fields. In particular, the gauge invariant fermion fields
can be constructed in terms of the exponential of Schwinger
line-integrals as follows,
\bear\label{2.14}
&&\hat\psi^{i\alpha}(x)= e^{-i\int_x^\infty dz^\mu a_\mu}
\left(Pe^{-i\int^\infty_x dz^\mu  A_\mu}\right)_{ij}
\left( Pe^{-i\int^\infty_x dz^\mu B_\mu}\right)_{\alpha\beta}
\psi^{j\beta}(x)\nonumber\\
&&\hat\chi^i(x)=e^{-i\int^\infty_x dz^\mu b_\mu}
\left( Pe^{-i\int^\infty_x dz^\mu A_\mu}\right)_{ij}\chi^j(x)
\ear
where $P$ denotes ``path-ordering''.

In section 5 we shall show that some primaries of the FQS models
with conformal dimensions given by the Kac formula (\ref{1.1b})
can be  constructed as suitable normal ordered products
of these fields. Furthermore, as we shall demonstrate for the
IM in section 4, the Schwinger line integrals in (2.14) play an
essential role in the realization of the order-disorder algebra.

\section{Reduction Formulae}
\setcounter{equation}{0}
In section 4 we shall be concerned with the calculation of four-point
functions of gauge-invariant local fermion bilinears such as
\be\label{3.1}
\Phi^{(k)}_{2,2}=\hat\psi_2^{\dagger i\alpha}\hat\psi_2^{i\alpha}+
\hat\psi_1^{\dagger i\alpha}
\hat\psi_1^{i\alpha}=\psi_2^{\dagger i\alpha}\psi_2^{i\alpha}+
\psi_1^{\dagger i\alpha}
\psi_1^{i\alpha}\ee
In terms of the decoupled fields, expression (\ref{3.1})
takes the form
\bear\label{3.2}
&&\Phi_{2,2}^{(k)}=\psi_2^{(0)\dagger i\alpha}
\tilde g_A^{ij}\tilde g_B
^{\alpha\beta}e^{2\phi_a}\psi_2^{(0)j\beta}\nonumber\\
&&+\psi_1^{(0)\dagger i\alpha}\left(\tilde g_A^{-1}\right)^{ij}
\left(\tilde g_B^{-1}\right)^{\alpha\beta}
e^{-2\phi_a}\psi_1^{(0)j\beta}
\ear
The fields are understood to be normal-ordered.
Their conformal dimensions are, respectively
\bear\label{3.3}
h_{\psi_\alpha}&=&\frac{1}{2}\nonumber\\
h_{e^{2\phi_1}}&=&h_{e^{-2\phi_1}}=-\frac{1}{4k}\ear
and \cite{KZ}
\bear\label{3.4}
&&h_{\tilde g_A}=h_{\tilde g_A^{-1}}=-\frac{3}{4(k+3)}\nonumber\\
&&h_{\tilde g_B}=h_{\tilde g_B^{-1}}=-\frac{(k^2-1)}{2k(k+2)}\ear
Note that the conformal dimensions (\ref{3.4})
correspond to WZW-fields
$\tilde g_A$ and $\tilde g_B$ in the fundamental representation of
$SU(2)_{-(k+5)}$ and $SU(k)_{-2(k+1)}$, respectively.

The dimension of the composite $\Phi^{(k)}_{2,2}$
is simply given in terms
of the sum of the individual contributions,
\be\label{3.5}
h_{2,2}^{(k)}=\frac{1}{2}-\frac{1}{4k}-\frac{3}{4(k+3)}
-\frac{(k^2-1)}{2k
(k+2)}=\frac{3}{4(k+2)(k+3)}\ee
with the same result for (antiholomorphic) dimension
$\bar h_{2,2}^{(k)}$.
The result (\ref{3.5}) agrees with Kac's
formula (\ref{1.1b}) for $p=q=2$. This
suggests the identification of $\Phi_{2,2}^{(k)}$
with a primary field.

A typical contribution to the four-point function
of $\Phi_{2,2}^{(k)}$ is
given by
\bear\label{3.6}
&&<(\psi^\dagger_2\psi_2)(1)(\psi^\dagger_1\psi_1)
(2)(\psi^\dagger_1\psi_1)(3)(\psi_2^\dagger\psi_2)(4) >=\\
&=&\frac{(\mu^2)^{\frac{1}{k}}}{16}
\left|\frac{z_{12}z_{13}z_{24}z_{34}}
{z_{14}z_{23}}\right|^{\frac{1}{k}}\left\lbrace
\frac{1}{|z_{12}z_{34}|^2}G_A(1,2,4,3)G_B(1,2,4,3)
\right.\nonumber\\
&&\qquad\qquad+\frac{1}{|z_{13}z_{24}|^2}G_A(1,3,4,2)
G_B(1,3,4,2)\nonumber\\
&&\qquad\qquad-\frac{1}{z_{12}z_{34}\bar z_{13}
\bar z_{24}}\hat G_A(1,3,4,2)
\hat G_B(1,3,4,2)\nonumber\\
&&\qquad\qquad\left.-\frac{1}{z_{13}z_{24}
\bar z_{12}\bar z_{34}}
\hat G_A(1,2,4,3)\hat G_B(1,2,4,3)\right\rbrace\nonumber\ear
where $z_{ij}=z_i-z_j,\bar z_{ij}=\bar z_i-\bar z_j$, and
\bear
G_A(1,2,3,4)&=&<tr (\tilde g_A(1)\tilde g_A^{-1}(2))
tr(\tilde g_A(3)\tilde g_A^{-1}(4))>\label{3.7}\\
\hat G_A(1,2,3,4)&=&<tr (\tilde g_A(1)\tilde g_A^{-1}(2)
\tilde g_A(3)\tilde g_A^{-1}(4))>\label{3.8}\ear
with the corresponding definition for $G_B$ and $\hat G_B$ in terms
of $\tilde g_B$. The kinematic factors within the curly brackets in
(\ref{3.6}) arise from the contractions of the free fermions,
as given by the two-point
functions
\bear\label{3.9}
<\psi^{(0)i_1}_1(1)\psi^{(0)\dagger i_2}_2(2)>&=&
\frac{1}{2}\frac{\delta^{i_1i_2}}
{z_{12}}\nonumber\\
<\psi^{(0)j_1}_2(1)\psi^{(0)\dagger j_2}_1(2)>&=&\frac{1}{2}
\frac{\delta^{j_1j_2}}
{\bar z_{12}}.\ear
and the overall factor multiplying (\ref{3.6}) arises
from the four-point
function of the exponentials $e^{\pm2\phi_1}$ upon using
\bear\label{3.10}
<e^{-\alpha\phi_1(1)}e^{\beta\phi_1(2)}>&=&|\mu z_{12}
|^{\frac{\alpha^2}{4k}},
\quad\alpha=\beta\nonumber\\
&=&0,\quad \alpha\not=\beta\ear
where $\mu$ is an arbitrary infrared regulator.

The functions (3.7) and (3.8) and the corresponding ones
for $\tilde g_B$
may directly be obtained from the work of ref. \cite{KZ}, by continuing
the KZ result to the
negative levels in question. Instead we shall make use of the duality
between $SU(N)_n$ and $SU(n)_{-(N+2n)}$ following from (\ref{2.3})
\cite{NS} to obtain a
``reduction formula'' which will prove very useful further on.

We begin by making the identifications
\bear\label{3.11}
&&g^{ij}=\mu^{-1}\psi_1^{\dagger j\alpha}\psi^{i\alpha}_1=\frac{1}{\mu}
\psi_1^{(0)\dagger_{j\alpha}}\tilde g_{\alpha\beta}e^{-2\phi}
\psi_1^{(0)i\beta}\nonumber\\
&&(g^{-1})^{ji}=\mu^{-1}\psi_2^{\dagger i\alpha}\psi^{j\alpha}_2=
\frac{1}{\mu}\psi_2^{(0)\dagger i\beta}
\tilde g_{\beta\alpha}^{-1}e^{2\phi}
\psi_2^{(0)j\alpha}\ear
where now
\be\label{3.12}
<e^{-\alpha\phi(1)}e^{\alpha\phi(2)}>=|\mu z_{12}
|^{\frac{\alpha^2}{2nN}}.\ee
Comparison of the respective four-point functions then leads to a
relation between the four-point function of $g\in SU(N)_n$ and $\tilde g
\in SU(n)_{-N-2n}$. With the identification (\ref{3.11}) we have
\bear\label{3.13}
&&<g(1)g^{-1}(2)g^{-1}(3)g(4)>=\left(\frac{1}{\mu}\right)^4
<e^{-2\phi(1)}e^{2\phi(2)}e^{2\phi(3)}e^{-2\phi(4)}>\\
&&\times<(\psi^{(0)\dagger}_1\tilde g\psi_1^{(0)})(1)
(\psi^{(0)\dagger}_2\tilde g^{-1}
\psi_2^{(0)})(2)(\psi^{(0)\dagger}_2\tilde g^{-1}\psi_2^{(0)})(3)
(\psi^{(0)\dagger}_1\tilde g\psi_1^{(0)})(4)>\nonumber\ear

\bear\label{3.14}
<g(1)g^{-1}(2)g^{-1}(3)g(4)>=&&\left(\frac{1}{\mu^2}
\right)^{2-\frac{2}{nN}}
\left|\frac{z_{12}z_{34}z_{13}z_{24}}{z_{143}z_{23}}
\right|^{\frac{2}{nN}}
\times\\
\times\frac{1}{16}\left\lbrace \right.&&I_1\bar I_1\frac{1}
{|z_{12}z_{34}|^2}<tr
(\tilde g(1)\tilde g^{-1}(2))tr(\tilde g^{-1}(3)\tilde g(4))>
\nonumber\\
+&&I_2\bar I_2\frac{1}{|z_{13}z_{24}|^2}<tr
(\tilde g(1)\tilde g^{-1}(3)tr(\tilde g^{-1}(2)\tilde g(4))>
\nonumber\\
-&&I_1\bar I_2\frac{1}{z_{13}z_{24}\bar z_{12}\bar z_{24}}<tr
(\tilde g(1)\tilde g^{-1}(3)\tilde g(4)\tilde g^{-1}(2))>\nonumber\\
-&&I_2\bar I_1 \left.\frac{1}{z_{12}z_{34}\bar z_{13}\bar z_{24}}<
tr(\tilde g(1)\tilde g(2)^{-1}\tilde g(4)\tilde g(3)^{-1})>\right\rbrace
\nonumber\ear
where $I_A,\bar I_B$ stand for the $SU(N)$ invariant tensors \cite{KZ}
\bear\label{3.15}
&&I_1=\delta^{i_1i_2}\delta^{i_3i_4},\quad \bar I_1=\delta^{j_1j_2}
\delta^{j_3
j_4}\nonumber\\
&&I_2=\delta^{i_1i_3}\delta^{i_2i_4},\quad \bar I_2=\delta^{j_1j_3}
\delta^{j_2
j_4}.\ear
Following \cite{KZ}, we make the decomposition
\be\label{3.16}
<g(1)g^{-1}(2)g^{-1}(3)g(4)>=|z_{14}z_{23}|^{-4h_g}
\sum_{A,B} I_A\bar I_B
G_{AB}(x,\bar x),\ee
where
\be\label{3.17}
h_g=\frac{N^2-1}{2N(N+n)},\ee
and $x$, $\bar x$ are given by
\be\label{3.18}
x=\frac{z_{12}z_{34}}{z_{14}z_{32}},\quad \bar x=\frac{\bar z_{12}
\bar z_{34}}
{\bar z_{14}\bar z_{32}}.\ee
The $G_{AB}(x,\bar x)$ are the positive level
WZW blocks which have been explicitly calculated
in ref. \cite{KZ}

Comparison  of both sides in eq. (\ref{3.10}) then yields
\bear\label{3.19}
&&<tr(\tilde g(1)\tilde g^{-1}(2)) \ tr(\tilde g^{-1}(4)\tilde g(3))>=
\\
&&=16|\mu^2z_{14}z_{23}|^{\frac{2(n^2-1)}{n(N+n)}}\frac{|x|^2}
{|x(1-x)|\frac{2}{nN}}G_{11}(x,\bar x)\nonumber\\
&&\nonumber\\
&&<tr(\tilde g(1)\tilde g^{-1}(2)\tilde g(4)\tilde g^{-1}(3)
)>=\nonumber\\
&&=-16|\mu^2z_{14}z_{23}|^{\frac{2(n^2-1)}{n(N+n)}}\frac{x(1-\bar x)}
{|x(1-x)|\frac{2}{nN}}G_{21}(x,\bar x)\nonumber\ear
The remaining vacuum expectation values are obtained via the
substitution $x\to1-x$. The fact that only two WZW blocks ($G_{11}$
and $G_{21}$) are needed is a consequence of the bosonic character of
the WZW field, which implies
\be\label{3.20}
G_{11}(1-x)=G_{22}(x),\ G_{12}(1-x)=G_{21}(x)\ee

As we shall see in the following section, relations (\ref{3.16})
will prove
very useful. The real power of this reduction
technique will, however,
come into play when considering the four-point
function of other primaries,
which, as we shall see in section 5, require the
calculation of four-point
correlators of WZW fields belonging to representations
of $SU(2)_{-(k+5)}$ and
$SU(k)_{-2(k+1)}$, other than fundamental ones. This
poses the problem that
the four-point correlator of $SU(k)$-WZW fields $(k\not=2$) are
only known for the fundamental representation. Via the reduction
technique described above one can relate the correlator in a given
representation of $SU(k)_{-2(k+1)}$ to a correlator
in the transposed
representation of $SU(2)_k$. Since the four-point correlators of
WZW fields have been calculated for any integrable
(spin $j$, $j=0,1,...,\frac{k}{2}$) representation $SU(2)_k$, this
allows one to calculate
the four-point correlators of $\tilde g_B$ in any
representation. As for
the corresponding correlators involving $\tilde g_A$ belonging to
$SU(2)_{-(k+5)}$, they may be obtained from known $SU(2)_k$
correlators for
positive level $k$, or from (\ref{3.16}) in the case of $k=1$ (Ising
model). In short, this means that the correlators of the
primaries ${\Phi}
^{(k)}_{p,q}$ in the fermionic coset representation of minimal models
can in principle be calculated for arbitrary values of $k$ in
terms of $SU(2)$-WZW four-point functions, modulo
free-field correlators,
in analogy with the bosonic coset case.

\section{Fermionic Coset Description of Ising Model}
\setcounter{equation}{0}

As is well known, the critical Ising model can be described by a
continuum field theory of massless, free Majorana fermions
$\psi_M$ and $\bar\psi_M$.
\cite{IZ},
\cite{BS}. In this description the energy density operator
$\epsilon(x)$
is given by the local operator product of two Majorana fields.
While this
representation has proven to be useful, a corresponding
local representation
for the order (spin) operator $\sigma(x)$ and its dual, the disorder
operator $\mu(x)$, is lacking.

In this section we use the ideas developed previously in order to
construct an explicit local representation of the operators
$\epsilon(x),
\sigma(x)$ and $\mu(x)$
as suitable products of gauged Dirac fermions,
of conformal dimensions 1/2, 1/16, and 1/16 respectively, in agreement
with the critical exponents of the Ising model. We also show that the
corresponding four-point functions agree with results
obtained by other
methods \cite{KC}, \cite{LP}, \cite{BPZ}. A corresponding
realization of the Majorana fermions is also given.

\vspace{1cm}
\noindent {\it a) Energy and order operators}

\vspace{.5cm}
As was mentioned in the introduction, the critical Ising model
corresponds to a
conformal field theory with central charge $c=1/2$, that is, to $k=1$
in the series (\ref{1.1b}). As seen from (\ref{1.1b}), the operators
$\varepsilon$ and $\sigma$ can thus be associated with the primary
fields $\Phi^{(1)}_{2,1}$ and $\Phi^{(1)}_{2,2}$ of conformal
dimension $1/2$ and $1/16$, respectively. In terms of our coset
description, this leads us to make the gauge invariant ansatz
\bear\label{4.1}
\Phi^{(1)}_{2,1}=\varepsilon&=&\frac{1}{\mu}\left[
\left(\hat\psi^{\dagger i}_2
\hat\chi^i_1+\hat\chi_2^{\dagger i}\hat\psi_1^i\right)
\left(\hat\chi_1^{\dagger i}\hat\psi^i_2+\hat\psi_1^{\dagger
i}\hat\chi^i_2\right)\right]\\
&=&-\frac{1}{\mu}\left(\psi_2^{(0)\dagger}:g_A^{-1}
g_A:\chi_1^{(0)}\right)\left(\chi_1^{(0)\dagger}:\bar g_A^{-1}\bar
g_A:\psi_2^{(0)}\right):e^{2\phi_a}::e^{-2\phi_b}:\nonumber\\
&&-\frac{1}{\mu}\left(\psi_2^{(0)\dagger}:g_A^{-1}
g_A:\chi_1^{(0)}\right)\left(\psi_1^{(0)\dagger}:  \bar g_A^{-1}\bar
g_A:\chi_2^{(0)}\right):e^{\varphi_a}::e^{-\bar\varphi_a}
:\nonumber\\
&&\qquad\qquad:e^{-\varphi_b}::e^{\bar\varphi_b}:\nonumber\\
&&+(\psi\leftrightarrow \chi),\varphi_a\leftrightarrow\varphi_b,
\bar\varphi_a\leftrightarrow\bar\varphi_b)\nonumber\ear
for the energy operator, and
\bear\label{4.2}
\Phi^{(1)}_{2,2}=\sigma&=&\frac{1}{\mu}\left(\hat\psi^{\dagger i}
\hat\psi^i+\hat\chi^{\dagger i}\hat\chi^i\right)\\
&=&\frac{1}{\mu}\left(\psi_2^{(0)\dagger}\tilde g_A
e^{2\phi_a}\psi_2^{(0)}+\psi_1^{(0)\dagger}e^{-2\phi_a}\tilde
g_A^{-1}\psi_1^{(0)}\right)+\left(\psi\to\chi,\phi_a,\to\phi_b\right)
\nonumber
\ear
for the order operator, where $\mu$ is a parameter of dimension
one, which we choose to coincide with the infrared regulator
in (3.9).
Recalling (3.2), and noting that $\exp(\pm 2\phi_b)$
has dimension $-1/4$, one checks that the two operators
$\varepsilon$ and $\sigma$ have the correct conformal
dimensions, once we make the identifications $:g_A^{-1}
g_A:=1$ and $:\bar g_A^{-1}\bar g_A:=1$ for the respective
normal ordered products.

For the fermionic coset formulation in question,
one has two different candidates for the $\Phi^{(1)}_{2,2}$
primary field.
The linear combination of bilinears in $\psi$ and $\chi$ in
(\ref{4.2}) is suggested by the known
operator product expansion of
$\sigma(x)\sigma(x+\varepsilon)$ \cite{BPZ,DFSZ}.
The specific form of $\epsilon$, on the other hand,
is suggested by the
usual identification of $\epsilon$ with the bilinear
$\psi_M\bar\psi_M$ of
Majorana spinors (see eqs. (4.28), (4.29)).

{}From (4.1) one sees that  all the multipoint correlation functions
of $\varepsilon$ can be calculated explicitly,
once we set $:g_A^{-1}g_A:=:\bar g_A^{-1}\bar g_A:=1$.
For the four-point function a straight-forward calculation yields
\footnote{Here $Pf$ denotes the ``Pfaffian'', defined in general by\\
$Pf(A_{ij})=\sum_P(-1)^P A_{i,j_1}...A_{i_nj_n}$, the sum being
taken over all possible permutations $P$. Henceforth we make use
of the arbitrariness of the parameter $\mu$, in order to normalize
our correlators appropriately.},
\bear\label{4.3}
\langle\varepsilon(1)\varepsilon(2)\varepsilon(3)
\varepsilon(4)\rangle&=&
\frac{1}{|z_{12}z_{34}|^2}\frac{|1-x+x^2|^2}{|1-x|^2}\nonumber\\
&=&\Bigl| Pf\left(\frac{1}{z_{ij}}\right)\Bigr|^2\ear
which agrees with the result obtained in the
Majorana formulation \cite{IZ}.

The evaluation of the four-point function of
$\sigma$ proceeds as in the
case of (\ref{3.4}). It is convenient to introduce the notation
\bear\label{4.4}
\alpha:&=&\psi^{\dagger i}_1\psi^i_1,\ \alpha^{-1}:=
\psi^{\dagger i}_2\psi^i_2\nonumber\\
\tilde\alpha:&=&\chi^{\dagger i}_1\chi_1^i,\ \tilde\alpha^{-1}:
=\chi^{\dagger i}_2\chi^i_2\ear
Taking account of the selection rules contained in eqs. (\ref{3.6})
and (\ref{3.7}), we are left with 16 terms  which, because of Bose
symmetry can each be reduced to the form
$\langle\alpha(1)\alpha^{-1}(2)\alpha^{-1}(3)\alpha(4)\rangle,\
\langle\tilde\alpha(1)\tilde\alpha^{-1}(2)\tilde\alpha^{-1}(3)
\tilde\alpha(4)\rangle$  or\\
$\langle\alpha(1)\alpha^{-1}(2)\tilde\alpha^{-1}
(3)\tilde\alpha(4)\rangle$, by suitable relabeling of the arguments.

The correlator $\langle\alpha(1)\alpha^{-1}(2)
\alpha^{-1}(3)\alpha(4)\rangle$
is given by (\ref{3.4}) with $k=1$ and $G_B=\hat G_B=1$,
corresponding to
the absence of the $B$-field in (2.5).
The  four-point functions $G_A$ and
$\hat G_A$ of the $SU(2)_{-6}$ WZW field are
conveniently calculated from
the reduction formulae (\ref{3.16}), with $N=n=2$.
In that case $G_{11}$
and $G_{21}$ are the WZW blocks of the $SU(2)_2$ field \cite{KZ}:
\bear\label{4.5}
G_{11}(x,\bar x)&=&\frac{|1-x|^{1/4}}{|x|^{3/4}}[F_1(x)F_1(\bar x)+
\frac{1}{4}|x|F_2(x)F_2(\bar x)]\nonumber\\
G_{21}(x,\bar x)&=&\frac{|1-x|^{1/4}}{|x|^{3/4}}
\left[\frac{1}{2}x F_3(x)F_1(\bar x)-\frac{1}{2}|x|
F_1(x)F_2(\bar x)\right].\ear
Here the $F_i$'s are the hypergeometric functions
\bear\label{4.6}
F_1(x)&=&F(\frac{1}{4},-\frac{1}{4};\frac{1}{2};x)=\frac{1}
{2}(f_1(x)+f_2(x))
\nonumber\\
F_2(x)&=&F(\frac{1}{4},\frac{3}{4};\frac{3}{2};x)=\frac{1}
{{\sqrt x}}(f_1(x)-f_2(x))\nonumber\\
F_3(x)&=&F(\frac{5}{4},\frac{3}{4};\frac{3}{2};x)=-\frac{1}
{\sqrt{x(1-x)}}(f_2(x)-f_1(x))\nonumber\\
F_4(x)&=&F(\frac{1}{4},\frac{3}{4};\frac{1}{2};x)=
\frac{1}{2}\frac{1}{\sqrt{1-x}}(f_1(x)+f_2(x))\ear
with
\be\label{4.7}
f_1(x)=\sqrt{1+{\sqrt x}},\ f_2(x)=\sqrt{1-\sqrt x}.\ee
Note that $f_1$ and $f_2$ are  solutions of the second order differential
equation arising from the null-vector condition for a
$\Phi^{(1)}_{2,2}$ field \cite{BPZ}.

{}From (4.5) and (\ref{3.16}) with $N=n=2$ we then obtain
\be\label{4.8}
G_A(1,2,4,3)=8|\mu^2z_{14}z_{23}|^{3/4}\frac{|x|}{|x(1-x)|^{1/4}}
\left(f_1(x)f_1(\bar x)+f_2(x)f_2(\bar x)\right)\ee
\be\label{4.9}
\hat G_A(1,2,4,3)=-8|\mu^2z_{14} z_{23}|^{3/4}\frac{\sqrt{x(1-\bar
x)}}{|x(1-x)|^{1/4}}(f_1(x)f_2(\bar x)-f_2(x)f_1(\bar x)).
\ee
Substitution of these results into (\ref{3.4}) (with $k=1$) then leads to
\be\label{4.10}
\langle\alpha(1)\alpha^{-1}(2)\alpha^{-1}(3)\alpha(4)\rangle=
\frac{1}{|\mu^2 z_{14}z_{23}|^{1/4}}\frac{1}{|x(1-x)|^{1/4}}
(f_1(x)f_1(\bar x)+f_2(x)f_2(\bar x)).\ee
It is remarkable that despite the appearance of two different
combinations of the $f_i$'s in (4.5) the final  result (\ref{4.7})
can be written after a number of manipulations in terms of the
first one of the combinations.

The same result is evidently obtained for
$\langle\tilde\alpha(1)\tilde\alpha^{-1}(2)
\tilde\alpha^{-1}(3)\tilde\alpha(4)
\rangle$. Note that expression (4.7) has the remarkable
property of being invariant under the permutation of the arguments.
Hence all unmixed correlators of the above type are given by
the r.h.s. of (\ref{4.7}). This leaves us with the calculation of
the mixed correlator  $\langle
\alpha(1)\alpha^{-1}(2)\tilde\alpha^{-1}(3)\tilde\alpha(4)\rangle$.
It is easy to see that this correlator only involves the trace (3.7).
Hence it involves the $f_i$'s only in  the combination (4.8).
In fact one finds
\be\label{4.11}
\frac{\langle\alpha(1)\alpha^{-1}(2)\tilde\alpha^{-1}(3)\tilde\alpha(4)
\rangle}{\langle\alpha(1)\alpha^{-1}(2)\alpha ^{-1}(3)\alpha(4)\rangle}=
\frac{1}{4}\ee
Adding all contributions we thus finally obtain for the four-point
correlator of the order operator
\bear\label{4.12}
&&\langle\sigma(1)\sigma(2)\sigma(3)\sigma(4)\rangle=\\
&&\frac{1}{|\mu^2 z_{14} z_{23}|^{1/4}}\frac{1}{|x(1-x)|^{1/4}}
\left(\sqrt{1+\sqrt x}\sqrt{1+\sqrt x}+\sqrt{1-\sqrt x}\sqrt{1-\sqrt{\bar
x}}\right)\nonumber\ear
where normalization constants have been absorbed into the arbitrary
mass scale $\mu$. This result agrees with the one obtained by \cite{BPZ}
using general conformal arguments.

\vspace{1cm}
\noindent{\it b) Disorder operator and dual algebra}

\vspace{.5cm}
A complete characterization of the Ising model must also include
the disorder operator $\mu$ of dimension $1/16$. This operator should
satisfy the equal-time dual algebra \cite{BS}
\be\label{4.13}
\sigma(1)\mu(2)=e^{i\pi\theta(x_1-x_2)}\mu(2)\sigma(1)\ee
where $x_i$ denotes the real part of $z_i$. This leads us to make the
following ansatz in terms of the gauge-invariant fermion fields (2.14):
\be\label{4.14}
\mu(x)=\hat\psi^\dagger_2\hat\chi_2+\hat\chi^\dagger_1\hat\psi_1+(\psi
\leftrightarrow\chi).\ee
The contribution of the non-abelian line integrals cancels as a
consequence of the underlying $SU(2)$-gauge invariance
of the bilinears,
as well as the
absence of singularities in $\langle \psi^{(0)^\dagger}_2
\chi_2^{\ (0)}\rangle$,
etc.  We are thus formally left with
\bear\label{4.15}
\mu(x)&=&\psi^\dagger_2 e^{i\int^\infty_x dz^\mu a_\mu}
e^{-i\int^\infty_x dz^\mu b_\mu}\chi_2+
+\chi^\dagger_1e^{i\int^\infty_x dz^\mu b_\mu}
e^{-i\int^\infty_x dz^\mu a_\mu}\psi_1\nonumber\\
&+& (\psi\leftrightarrow\chi,a_\mu
\leftrightarrow b_\mu).\ear
In terms of the decoupled fields (\ref{2.6}) we thus obtain
\bear\label{4.16}
\mu(x)&=&\left(\psi_2^{(0)\dagger}\tilde g_A\chi_2^{(0)}
\right)e^{\varphi_a}
e^{\bar\varphi_b}+\left(\chi^{(0)\dagger}_1\tilde g_A^{-1}
\psi_1^{(0)}\right)
e^{-\varphi_a}e^{-\bar\varphi_b}+\nonumber\\
&&+(\psi\leftrightarrow\chi,\varphi_a\leftrightarrow
\varphi_b,\bar\varphi
\leftrightarrow\bar\varphi_b)\ear
where $(i=a,b)$
\bear\label{4.17}
\varphi_i&=&\phi_i+i\int^\infty_x dz^\mu
\varepsilon_{\mu\nu}\partial_\nu\phi_i,
\nonumber\\
\bar\varphi_i&=&\phi_i-i\int^\infty_x
dz^\mu\varepsilon_{\mu\nu}\partial_\nu\phi_i,\ear
are respectively the holomorphic and anti-holomorphic
components of the fields $\phi_i$ parametrizing
$a_\mu$ and $b_\mu$. Using the (euclidean) equal-time commutator
\be\label{4.18}
[\phi_i(x),\partial_0\phi_i(y)]_{ET}=-\frac{\pi}{2}\delta(x^1-y^1)\ee
we have from (\ref{4.17})
\bear\label{4.19}
&&[\varphi_i(x_1),\varphi_i(x_2)]_{ET}=\frac{-i\pi}{2}
\epsilon(x_1-x_2)
\nonumber\\
&&[\bar\varphi_i(x_1),\bar\varphi_i(x_2)]_{ET}=
\frac{-i\pi}{2}\epsilon(x_1-x_2)
\ear
\be\label{4.20}
[\varphi_i(x_1),\bar\varphi_i(x_2)]_{ET}=\frac{i\pi}{2}\ee
Making use of these commutation relations we obtain from (\ref{4.2})
and (\ref{4.16}) the equal-time duality relation (\ref{4.13}), as
required.

The evaluation of the 4-point function of the $\mu$-operator
proceeds along
the same lines as in the case of the 4-point function of the order
operator. The result is again given by the r.h.s. of
(\ref{4.12}), as
expected \cite{KC}.

We next calculate the mixed 4-point correlation function
$<\sigma(1)\mu(2)\sigma
(3)\mu
\newline
(4)>$. To this end we  introduce in
addition to (4.4) the notation
\bear\label{4.21}
\beta=\hat \psi_1^\dagger\hat\chi_1&&\quad\beta^{-1}=
\hat\chi_2^\dagger\hat\psi_2
\nonumber\\
\tilde\beta=\hat\chi_1^\dagger\hat\psi_1&&\quad\tilde\beta^{-1}=
\hat\psi_2^\dagger\hat\chi_2\ear
In the mixed case the evaluation is less straightforward than in the
case of the unmixed four-point functions, since the operators $\sigma$
and $\mu$ no longer commute.

The fermionic selection rules lead us to consider the evaluation
of 24 terms of the type $<\alpha(1)\beta(2)\alpha^{-1}
(3)\beta^{-1}(4)>,
<\alpha(1)\alpha^{-1}(2)\tilde\beta(3)\tilde\beta^{-1}(4)>,\\
 <\alpha(1)\beta^{-1}(2)\tilde\alpha(3)\tilde\beta^{-1}(4)>,\
<\alpha^{-1}(1)\tilde\alpha^{-1}(2)\beta(3)\tilde\beta(4)>,$
as well as those resulting from the permutation of
the arguments, and the
exchange $\gamma\leftrightarrow\tilde \gamma$, where $\gamma$
stands generically
for the operators (\ref{4.4}) and (\ref{4.21}). This number can
be reduced by noting that two correlators related by the interchange
 $\gamma\leftrightarrow\tilde \gamma$ are equal.
We outline the calculation
for the case of two typical terms.

\bigskip
i) Consider $<\alpha(1)\beta(2)\alpha^{-1}(3)\beta^{-1}(4)>$. From
(2.6)  we have
\bear\label{4.22}
&&<\alpha(1)\beta(2)\alpha^{-1}(3)\beta^{-1}(4)>=<e^{-(\varphi_a+
\bar\varphi_a)(1)}e^{-(\varphi_b+\bar\varphi_a)(2)}e^{(\varphi_a+
\bar\varphi_a)(3)}e^{(\varphi_b+\bar\varphi_a)(4)}>\nonumber\\
&&\\
&&\times <(\psi_1^{(0)\dagger}\tilde g_A\psi_1^{(0)}(1)
(\psi^{(0)\dagger}_1\tilde g_A\chi_1^{(0)})(2)(\psi_2^{(0)\dagger})
 \tilde g_A^{-1}\psi_2^{(0)})(3)
(\chi_2^{(0)\dagger}\tilde g_A^{-1}\psi_2^{(0)})(4)>\nonumber\ear
where normal ordering with respect to the free bosons and
fermions is understood. Recalling (\ref{3.9}) and (\ref{3.10}) we find
\bear\label{4.23}
&&\langle \alpha(1)\beta(2)\alpha^{-1}(3)\beta^{-1}(4)\rangle
=\mu^2\frac{e^{i\frac{\pi}{4}\eta}}{16}\left(\frac{z_{13}z_{24}\bar z
_{13}\bar z_{24}\bar z_{14}\bar z_{23}}{\bar z_{12}\bar z_{34}}
\right)^{1/2}\cdot\\
&&\cdot\left(\frac{1}{z_{13}z_{24}\bar z_{13}\bar z_{24}}G_A(1,3,2,4)
-\frac{1}
{z_{13}z_{24}\bar z_{14}\bar z_{23}}\hat G_A(1,3,2,4)
\nonumber\right)\ear
where the phase $\exp(i\pi\eta/4)$ arises from the
commutation relations
(\ref{4.20}). For the case in question $\eta=2$;
in general it takes the
values $\pm2$. Evaluation of (\ref{4.23}) shows that
it reduces, after a
number of manipulations\footnote{In particular one
makes use of $sgn (Im x)\frac{1}{|\mu^2 z_{14}z_{23}|^{1/4}}
\frac{1}{|x(1-x)|^{1/4}}(f_1(\dot x)f_1(\bar x)-f_2(x)f_2(\bar x))
=\frac{i}{|\mu^2z_{13}z_{24}|^{1/4}}\frac{1}{|y(1-y)|^{1/4}}
(f_1(y)f_2(\bar y)-f_2(y)f_1(\bar y))$ where $Im x$ stands for
``imaginary part of $x$'', and $y=x/(x-1)$.} to the
remarkably simple
result $(y=x/(x-1))$,
\be\label{4.24}
<\alpha(1)\beta(2)\alpha^{-1}(3)\beta^{-1}(4)>=\frac{i}{
|\mu^2 z_{13}z_{24}|^{1/4}}\frac{1}{|y(1-y)|^{1/4}}
(f_1(y)f_2(\bar y)-f_2(y)f_1(\bar y))\ee
this time involving another combination of the $f_i$'s in accordance
with the expected analiticity properties of the result.
(Compare with (4.8)). A numerical factor has again been
absorbed into the arbitrary
parameter $\mu$.

\bigskip
ii) We next consider $<\alpha(1)\beta^{-1}(2)\tilde\alpha(3)
\tilde\beta^{-1}(4)>$. One has this time
\bear\label{4.25}
&&<\alpha(1)\beta^{-1}(2)\tilde\alpha(3)\tilde\beta^{-1}(4)>\nonumber\\
&&=<e^{-(\varphi_a(1)+
\bar\varphi_a(1))}e^{(\varphi_b(2)+\bar\varphi_a(2))}
(e^{-(\varphi_b(3)+
\bar\varphi_b(3))}e^{(\varphi_a(4)+\bar\varphi_b(4))}>
\nonumber\\
&&\times <\left(\psi_1^{(0)\dagger}\tilde
g_A\psi_1^{(0)}\right)(1)\left(\chi^{(0)\dagger}_2\tilde
g_A^{-1}\psi_2^{(0)}\right)(2)\left(\chi_1^{(0)\dagger}
\tilde g_A\chi_1^{(0)}\right)(3)
\left(\psi_2^{(0)\dagger}\tilde g_A^{-1}\chi_2^{(0)}\right)(4)>
\nonumber\\
&&=\frac{1}{16}e^{\frac{i\pi\eta}{4}}(z_{14}z_{23}\bar z_{12}\bar
z_{34})^{1/2}\frac{1}{z_{14}z_{23}\bar z_{12}\bar z_{34}}
\hat G_A(1,4,3,2)\ear
The phase in this case corresponds to $\eta=-2$.
Expression (\ref{4.25})
is seen to involve the $f_i$'s only in the combination of
(\ref{4.9}). In
fact one obtains
\be\label{4.26}
\frac{<\alpha(1)\beta^{-1}(2)\tilde\alpha(3)\tilde\beta^{-1}(4)>}
{<\alpha(1)\beta(2)\alpha^{-1}(3)\beta^{-1}(4)>}=-1\ee
Explicit calculation shows that up to  numerical
factors, the same results are obtained for the remaining
terms contributing
to the mixed correlator. Absorbing again a normalization
constant into
the arbitrary scale parameter $\mu$, we finally have for the mixed
correlator $(y=x/(x-1))$
\be\label{4.27}
<\sigma(1)\mu(2)\sigma(3)\mu(4)>=\frac{i}{|\mu^2z_{14}z_{23}|^{1/4}}
\frac{1}{|y(1-y)|^{1/4}}\left(f_1(y)f_2(\bar y)-
f_2(y)f_1(\bar y)\right)\ee
This result agrees with the one obtained by BPZ using general
conformal arguments\footnote{There is a misprint in the relative
sign of the result of BPZ (eq. (I.39) of \cite{BPZ}).}.
 \cite{BPZ}. This provides further support for our
ans\"atze (\ref{4.1}), (\ref{4.2}) and (\ref{4.14}).

\vspace{1cm}
\noindent{\it c) Realization of Onsager fermions}

\vspace{.5cm}
To complete our discussion of the Ising model, we give a realization
of the Onsager fermions $\psi_M(x)$ and $\bar\psi_M(x)$ \cite{On}
in the fermionic
coset framework. We identify these fermions with the gauge-invariant
composites
\bear\label{4.28}
\psi_M&=&\hat\psi^\dagger_2\hat\chi_1+\hat\chi_2^\dagger\hat\psi_1
\nonumber\\
&=&\frac{1}{\sqrt\mu}\left(\psi_2^{(0)\dagger}
\chi_1^{(0)}:e^{-\varphi_a}
e^{\varphi_b}+\chi_2^{(0)\dagger}
\psi_1^{(0)}:e^{-\varphi_a}e^{\varphi_b}:\right)\ear
and
\bear\label{4.29}
\bar\psi_M&=&\hat\psi^\dagger_1\hat\chi_2+\hat\chi_1^\dagger\hat\psi_2
\nonumber\\
&=&\frac{1}{\sqrt\mu}\left(\psi_1^{(0)\dagger}\chi_2^{(0)}:e^{-\bar
\varphi_a}e^{\bar\varphi_b}:+\chi_1^{(0)\dagger}
\psi_2^{(0)}:e^{\bar\varphi_a}e^{-\varphi_b}:\right)\ear
of dimensions $(\frac{1}{2},0)$ and $(0,\frac{1}{2})$,
respectively. This
assignment agrees with the usual representation of the energy operator
(\ref{4.1}) in terms of Majorana fermions, $\epsilon=\psi_M\bar\psi_M$
\cite{IZ}. Note that $\psi^\dagger_M=\bar\psi_M$, as required.

\section{Other statistical models in the minimal unitary series}
\setcounter{equation}{0}
Our fermionic coset formulation is expected to allow
for the realization of all the primaries in the FQS series as
products of the fundamental, gauge-invariant fields (2.14).
Let us illustrate
this for the primaries $\Phi^{(k)}_{p,p}$ and $\Phi^{(k)}_{p,p-1}$,
the conformal dimension of a general primary $\Phi^{(k)}_{p,q}$
being given by (1.2).

We take the bosonic representation in terms of
gauged WZW fields \cite{GK}
as a guideline for our construction. In ref.
\cite{GK} the primaries in
the FQS series are constructed in accordance with (2.2) in terms of
SU(2) WZW fields $g$ and $g'$ of level $k$ and 1, respectively,
in a gauged WZW theory.
This motivates us to consider as basic building blocks the bilinears
\cite{NS}
\bear\label{5.1}
\hat g^{ij}&=&:\hat\psi_2^{i\alpha}\hat\psi_2^{\dagger j\alpha}:,
\quad \hat {g'}^{ij}=:\hat\chi_2^i\hat\chi^{\dagger j}_2:\nonumber\\
(\hat g^{-1})^{ij}&=&:\hat\psi_1^{i\alpha}\hat
\psi_1^{\dagger j\alpha}:, \quad
(\hat {g'}^{-1})^{ij}=:\hat\chi_1^i\hat\chi_1^{\dagger j}:\ear
Note that expressions (\ref{5.1}) involve the gauge-invariant fields
(2.14), so that $\hat g^{ij}$ and $(\hat g^{-1})^{ij}$
should be identified
with the gauge-invariant WZW fields of the gauged WZW action referred
to above.

According to ref. \cite{GK} we need to consider fields $\hat g_j$ in
the isospin $j$ representation of $SU(2)_k$ $(j=0,\frac{1}{2},1,..
\frac{k}{2})$. They can be constructed as the symmetrized direct
product of $2j$ fundamental fields (\ref{5.1}):
\be\label{5.2}
\hat g_j=\left[:\underbrace{\hat g\otimes...\otimes\hat g:}_{2j\
{\rm times}}
\right]_{\cal S}\ee
where ${\cal S}$ stands for the symmetrization with respect to left
and right
indices separately. We again restrict ourselves to the
isospin zero sector.
In this
sector we have for $\Phi^{(k)}_{p,p}$
\be\label{5.3}
\Phi^{(k)}_{p,p}=Tr(\hat g_j+\hat g_j^{-1})=Tr(g_j+g_j^{-1})\ee
where $p=2j+1$. The second equality in (\ref{5.3}) is a result of
the cancellation of the Schwinger line integrals upon
taking the trace.
In the notation of (2.4) the primary (\ref{5.3}) is given by
\bear\label{5.4}
\Phi^{(k)}_{p, p}&=&g_j^{(0)^{i_1...i_r,k_1...k_r}}.
\left(\tilde g_A^{-1}\right)_j^{k_1...k_r,i_1...i_r} \nonumber\\
&&+\left(g^{(0)}_j\to g^{(0)-1}_j,\
\tilde g_A\to\tilde g_A^{-1}
\right)\ear
where
\bear\label{5.5}
g^{(0)i_1...i_r,k_1...k_r}_j&=&
:e^{2r\phi_1}::\psi_2^{(0)\dagger k_1\gamma_1}...
\psi_2^{(0)\dagger k_\gamma\alpha_\gamma}::
\psi_2^{(0)^{i_1\alpha_1}}...\psi_2^{(0)i_r\alpha_r}:\nonumber\\
&&\cdot[:\tilde g^{\gamma_1\alpha_1}_B...
\tilde g^{\gamma_r\alpha_r}_B:]_{\cal A}
\nonumber\\
\left(\tilde g_A\right)_j^{k_1..k_r, i_1...i_r}&=&
\left[:\tilde g_A^{k_ji_1}...\tilde g_A^{k_ri_r}:\right]
_{\cal S}\ear
with $r=2j$, and where the superscript ``$(0)$'' in the l.h.s. means
that the diagonal
$SU(2)_{k+1}$ gauge field has been decoupled.
The field $g_j^{(0)}$ in
(\ref{5.5}) is the fermionic coset representation
of the $SU(2)_k$ WZW
field in the $j$-representation \cite{NS}.

Due to the Pauli principle the direct product of
$\tilde g_A$'s and
$\tilde g_B$'s project into the $(2j+1)$-dimensional symmetric
(${\cal S}$),
antisymmetric (${\cal A}$)
representation of $SU(2)$ and $SU(k)$, respectively,
with the corresponding
Casimirs given by \cite{NS}
\bear\label{5.6}
C({\cal S})&=&j(j+1)\nonumber\\
C({\cal A})&=&(k+2)\left[j-\frac{j^2}{k}-\frac{j(j+1)}{k+2}\right]
\ear
In terms of these Casimirs, we have for the conformal dimension of the
corresponding products
\bear\label{5.7}
h_{(\tilde g_A)_j}&=&\bar h_{(\tilde g_A)_j}=\frac{-j(j+1)}{k+3}
\nonumber\\
h_{\cal A}&=&\bar h_{\cal A}
=-\frac{C({\cal A})}{k+2}=-\left[j-\frac{j^2}{k}-\frac{j(j+1)}{k+2}
\right]\ear
Adding to $h_{\cal A}$ the contributions from the vertex
operator $(h_{e^{2r
\phi_1}}=\bar h_{e^{2r\phi_1}}=-\frac{j^2}{k})$ and the
free fermions, we
obtain
\be\label{5.8}
h_{g_j(0)}=\frac{j(j+1)}{k+2}\ee
We thus obtain $(p=2j+1)$
\be\label{5.9}
h_{p,p}^{(k)}=h_{g_j^{(0)}}+h_{(\tilde g_A)_j}=
\frac{j(j+1)}{(k+2)(k+3)}
=\bar h_{p,p}\ee
in agreement with the Kac formula (\ref{1.1b}).
On the other hand, in terms of the bilinears in $\chi_\alpha^i$, eq.
(\ref{5.2}), the Pauli principle only allows us
to construct the primary
\bear\label{5.10}
\Phi^{(k)}_{k+1,k+1}&=&Tr(\hat {g'}+\hat {g'}^{-1})\nonumber\\
&=&Tr\left({g'}_{\frac{1}{2}}^{(0)}\tilde g_A+{g'}^{(0)-1}
_{\frac{1}{2}}
\tilde g_A^{-1}\right)\ear
where
\[{g'}^{(0)ij}_{\frac{1}{2}}=:e^{2\phi_2}\chi_2^{(0)i}
\chi_2^{(0)\dagger j}:\]
with dimension (see eq. (3.2))
\be\label{5.11}
h_{k+1,k+1}^{(k)}=\bar h^{(k)}_{k+1,k+1}=\frac{k}{4(k+3)},\ee
again in agreement with Kac's formula.
Note that the primary (\ref{5.10})
has the same dimension as the primary (\ref{5.4}) for the maximum
value of $j,j=\frac{k}{2}$. Note also that for $k=1$, these primaries
are just the components making up the order operator (\ref{4.1}) in
the Ising model.

Following again ref. \cite{GK}, we  take for $\Phi^{(k)}_{p,p-1}$,
\be\label{5.12}
\Phi^{(k)}_{p,p-1}={\rm Tr}\left\{\hat g_j(\hat g'\otimes
 {1\!{\rm l}}_{j
-\frac{1}{2}})^{-1}+(\hat g'\otimes {1\!{\rm l}}
_{j-\frac{1}{2}})\hat g^{-1}_j
\right\},
\ee
where, in terms of the bilinears (\ref{5.1}),
\bear\label{5.13}
&&{\rm Tr} \left\{\hat g_j\left(\hat g'\otimes
 {1\!{\rm l}}_{j-\frac{1}{2}}
\right)^{-1}\right\}=\nonumber\\
&&\left\{ \left[\left( \hat\psi_2^\dagger\hat\psi_2\right)^{i_1j_1}
\ldots\left(\hat\psi_2^\dagger\hat\psi_2\right)^{i_rj_r}\right]_{\cal S}
\left(
\hat\chi_1^{\dagger j_1}\hat\chi_1^{i_1}\delta^{j_2i_2}\ldots
\delta^{j_ri_r}\right)\right\},
\ear
where ${\cal S}$ again denotes symmetrization  with
respect to the indices
$\{i_l\}$ and $\{j_l\}$,  separately. Expressing the fermionic fields
in terms of the decoupled variables (\ref{2.6}), one  finds,
\bear\label{5.14}
\Phi^{(k)}_{p,p-1}&=&:e^{2r\phi_1}::e^{-2\phi_2}::
\psi_2^{(0)^\dagger  k_1\beta_1}\ldots \psi_2^{(0)^\dagger j_r
\beta_r}:\nonumber\\
&\cdot&:\psi_2^{(0)j_1\alpha_1}\ldots\psi_2^{(0)j_r\alpha_r}:\left[
:\tilde g_B^{\beta_1 \alpha_1}\ldots
\tilde g_B^{\beta_r\alpha_r}:\right]
_{\cal A}\nonumber\\
&\cdot&\left[\tilde g_A^{k_2i_2}\ldots\tilde g_A^{k_ri_r}\right]_{\cal S}:
\chi_1^{(0)^\dagger j_1}\chi_1^{(0)k_1}:\delta^{j_2i_2}
\ldots\delta^{j_ri_r}.\ear
Note that we are left with the symmetrization of only
$r-1$ fields $g_A$.
This is a combined effect of the cancellation of the
Schwinger line integrals
and the use of $g_1g_1^{-1}=1,\bar g_1\bar g_1^{-1}=1$
within the normal
product. Using (3.3), (3.4) and (\ref{5.7}) one readily
checks that the
primary (\ref{5.12}) has the conformal dimension $(r=2j)$
\bear\label{5.15}
h^{(k)}_{p,p-1}&=&-\frac{1}{4}-\frac{r^2}{4k}+\frac{1}{2}(r+1)
-\left[ \frac{r}{2}-\frac{r^2}{4k}-\frac{r(r+2)}{4(k+2)}\right]-
\frac{(r^2-1)}{4(k+3)}\nonumber\\
&&=\frac{(r+k+3)^2-1}{4(k+2)(k+3)}\ear
which is in agreement with Kac formula (\ref{1.1b}).

\vspace{1cm}
\noindent{\it Other primaries}

\vspace{.5cm}
The primaries considered so far were constructed as local products of
chiral bilinears of the gauge-invariant fermion fields (\ref{2.14}).
The construction of the remaining primaries will be
carried out in the decoupled picture, where the requirement of BRST
invariance is more easily realized.

 It follows from the Lagrangian
(\ref{2.6}) that we have the BRST charges $Q_{U(1)}^{(a)},\
Q^{(b)}_{U(1)},\
Q^{(B)}_{SU(k)}$ and $Q^{(A)}_{SU(2)}$, associated with the gauge
fields $a_\mu,\ b_\mu,\ B_\mu$ and $A_\mu$, respectively.
Our construction will involve bilinears commuting with the first three
charges. Hence we need to consider only $Q:=Q^{(A)}_{SU(2)}$.

Following ref. \cite{He}, we have for the BRST charge $Q$,
\be\label{5.16}
Q=\oint dz:\left[\eta^a(z)G^a(z)+\frac{1}{2} C_{abc}\eta^a\eta^b
{\cal P}^c\right]:\ee
where $G^a(z)\approx 0$ are the first-class constraints
associated with
the diagonal $SU(2)$ gauge symmetry, $C_{abc}$ are the
structure constants of the corresponding constraint algebra,
\be\label{5.17}
\left\{G^a(z),G^b(w)\right\}=C_{abc}G^c(z)\delta(z-w)\ee
and
\bear\label{5.18}
&&\{{\cal P}^a,\eta^b\}=-\delta^{ab}\nonumber\\
&&(\eta^a)^*=\eta^a,\ ({\cal P}^a)^*=-{\cal P}^a\ear
In order to deduce the first-class constraints $G^a\approx 0$ from the
decoupled  partition function (2.9), we follow refs. \cite{P},
\cite{KS} by simultaneously gauging the sectors corresponding to
the free fermions (both $\psi$'s and $\chi$'s), $SU(2)_{-(k+5)}$
WZW field $\tilde g_A$, and $SU(2)$-ghost fields with an
external $SU(2)$ gauge field $W^a_\mu$. In terms of the variables
$w$ and $\bar w$ defined by
\be\label{5.19}
W=i(\bar\partial w)w^{-1},\quad\bar W=i(\partial \bar w)
\bar w^{-1}\ee
the dependence of the different sectors on the external gauge field
factors are as follows:
\bear\label{5.20}
&&\int{\cal D}\psi{\cal D}\psi^\dagger\int{\cal D}
\chi{\cal D}\chi^\dagger
exp\{-\frac{1}{\sqrt2\pi}\int\psi^\dagger({\slp}
+i{\slW})\psi\}exp\{-\frac{1}{\sqrt2\pi}
\int\chi({\slp}+i{\slW})\chi\}\nonumber\\
&&\quad\qquad\qquad\qquad\qquad =Z_F\ e^{(k+1)W[w^{-1}w]}\nonumber\\
&&\nonumber\\
&&\int{\cal D}\eta{\cal D}{\cal P}\int{\cal D}
\bar\eta{\cal D}\bar{\cal P}
e^{-\int{\cal P}^a\bar{\cal D}^{ab}(w)\eta^b}e^{-\int\bar{\cal P}^a
{\cal D}^{ab}(\bar w) \bar\eta^b}\nonumber\\
&&\quad\qquad\qquad\qquad\qquad =e^{2C_VW[w^{-1}\bar w]}
\int{\cal D}\eta{\cal D}{\cal P}\int{\cal D}\bar\eta{\cal D}
\bar{\cal P}
e^{-\int{\cal P}^a\bar\partial\eta^a}e^{-\int\bar{\cal P}^a\partial
\bar\eta}\nonumber\\
&&\nonumber\\
&&\int{\cal D}\tilde g_A e^{(k+5)\tilde W[\tilde g_A,w,\bar w]}=
e^{-(k+5)W[w^{-1}\bar w]}
\int {\cal D}\tilde g_A
e^{(k+5)\tilde W[\tilde g_A]} \ear
where $\tilde W[\tilde g_A,w]$ stands for the gauged WZW action
\be\label{5.21}
\tilde W[\tilde g_A,w,\bar w]=W[w^{-1}\tilde g_A\bar w]
-W[w^{-1}\bar w]\ee
and use has been made of the invariance of the Haar measure.
Note that for the case in question $C_V=2$, so that the dependence of
the corresponding gauged partition function
 $Z[w,\bar w]$ on the external field is seen to cancel, implying
\be\label{5.22}
\frac{i^n}{Z[w]}\frac{\delta^n\ Z[w]}{\delta w^{a_1}(z_1)\ldots\delta
w^{a_n}(z_n)}\Bigm\vert_{w=0}=\langle J^{a_1}(z_1)\ldots J^{a_n}(z_n)
\rangle=0\ee
where
\be\label{5.23}
J^a(z)=\frac{1}{i}\frac{\delta}{\delta w^a(z)} S[w,\bar w]
\Bigm\vert_{w=0}\ee
with $S[w,\bar w]$ the corresponding gauged action.
We are thus led to make the identification
\be\label{5.24}
G^a(z)=J^a(z)=j^a_\psi(z)+j^a_\chi(z)+\tilde j^a(z)+j^a_{gh}(z)\ee
where
\bear\label{5.25}
j^a_\psi&=&\frac{1}{\sqrt2\pi}\psi_2^{(0)^\dagger}
t^a\psi_1^{(0)}\nonumber\\
j^a_\chi&=&\frac{1}{\sqrt2
\pi}\chi_2^{(0)^\dagger}t^a\chi_1^{(0)}\nonumber\\
\tilde j^a(z)&=&-\left(\frac{k+5}{2}\right)tr\left( t^a\tilde g_A
\partial\tilde g^{-1}_A\right)\nonumber\\
j^a_{gh}(z)&=&f^{abc}{\cal P}^b\eta^c.\ear
Analogous relations apply to the antiholomorphic part.

Alternatively, the BRST charge (\ref{5.14}) may also be obtained as the
Noether
 charge associated with the invariance of the effective action in the
decoupled
picture \cite{Ta,Ba}.

The primaries $\Phi_{p,q}^{(k)}$ are constructed subject to the
requirement
that they commute with the BRST charge (\ref{5.16}).
For $p-q\geq 2$ this
construction will also have to include the currents (\ref{5.25})
as we shall
 see below. Our starting point is the observation that
\be\label{5.27}
h_{g_j^{(0)}}+h_{(\tilde g_A)_\ell}+\frac{(1-(-1)^{p-q})}{2}
h_{g^{\prime(0)}_{
\frac{1}{2}}}+N=h_{p,q}^{(k)}\ee
where $j=(p-1)/2$ and $\ell=(q-1)/2$,
and $N$ is an {\it integer} given by
\be\label{5.28}
N=\frac{1}{4}\left[(p-q)^2-\frac{(1-(-1)^{p-q})}{2}\right]\ee

Eq. (\ref{5.27}) tells us that we can obtain the primaries
by taking Kac-Moody
descendants of level $N$, from composites $\phi_h$ of the fields
$g_j^{(0)},(\tilde g_A)_\ell$ and $g^{\prime(0)}_{\frac{1}{2}}$
(if $p-q$ is odd),
using the property
\be\label{5.29}
h_{{\cal J}_{-N}\phi_h}=h+N\ee
where
\be\label{5.30}
{\cal J}_{-N}^a\phi_h(z)=\oint_{e_z}\frac{d\zeta}{2\pi i}\frac{{\cal
J}^a(\zeta)\phi_h(z)}
{(\zeta-z)^N}\ee
with
\be\label{5.31}
{\cal J}^a=j^a_\psi+\alpha j^a_\chi+\beta\tilde j^a.\ee
The coefficients $\alpha$ and $\beta$ are then fixed by
the  requirement of
 BRST invariance of the primary (\ref{5.28}).

We shall illustrate the procedure for the case $p-q=2,$ that is $N=1$.
We restrict ourselves to the isospin-zero sector. Invariance under
$SU(2)_L\times
SU(2)_R$ leads one to consider
\bear\label{5.32}
\Phi^{(k)}_{p,p-2}&=&tr :({\cal J} g^{(0)}_j\bar{\cal J}(\tilde
g^{-1}_A)_{j-1}): \ +c.c.
\nonumber\\
&\equiv&:\epsilon_{i_1\ell_1}\delta_{i_2\ell_2}{\cal J}_{\ell_1\ell
_2}(g^{(0)}_j)^{i_1i_2i_3
\ldots i_{2j},k_1k_2k_3\ldots k_{2j}}\nonumber\\
&&\cdot\delta_{k_1 n_1}\epsilon_{k_2n_2}\bar J_{n_1n_2}
(\tilde g_A^{-1})_{j-1}
^{k_3\ldots k_{2j},i_3\ldots i_{2j}}:
\ear
where ${\cal J}={\cal J}^at^a$. We now require
\be\label{5.33}
Q\Phi^{(k)}_{p,p-2}(w,\bar w)|\Omega\rangle=
\left[ Q,\Phi^{(k)}_{p,p-2}
(w,\bar w)\right]|\Omega\rangle=0\ee
where $|\Omega\rangle$ denotes the ground state.
This requires that the
short-distance expansion of $J^a(z)\Phi^{(k)}
_{p,p-2}(w,\bar w)$  be regular. However,
 because of the normal product appearing in (\ref{5.32}),
the computation of the
 commutator is more easily done by expanding $Q$
in terms of modes. From
 (\ref{5.30}) we have
\be\label{5.34}
:{\cal J}^a(z)\phi_h(z):={\cal J}^a_{-1}\phi_h(z).\ee
With the mode expansion
\be\label{5.35}
\eta^a(z)=\sum^\infty_{n=-\infty}z^n\eta^a_{-n},
J^a(z)=\sum^\infty_{n=-\infty}
z^n {\cal J}^a_{-n}\ee
we have from eq. (5.16), (5.25), (5.30) and (5.31),
\be\label{5.36}
\left[ Q,\Phi^{(k)}_{p,p-2}\right]|\Omega\rangle=\sum_{n=0}\
\eta^a_{-n}
\left[J^a_n,{\cal J}^b_{-1}\Phi^b\right]|\Omega\rangle=0\ee
where
\be\label{5.37}
\Phi^b:=tr\left(t^b g_j^{(0)}\bar {\cal J}_{-1}(\tilde
g_A^{-1})_{j-1}\right)+c.c.\ee
and use thas been made of $\eta^a_{-n}|\Omega\rangle=0$, for $n<0$.
Associativity,
 as well as the commutation relations
\be\label{5.38}
[j^a_n,j^b_m]=f_{abc}j^c_{n+m}+\frac{K}{2} n\delta^{ab}
\delta_{n,-m}\ee
for the currents (5.25) with level $K=k,1,-(k+5)$ and $2C_V=4$,
respectively,
gives
\be\label{5.39}
\left[J_n^a,{\cal J}^b_{-1}\Phi^b\right]=\left[ f_{abc}{\cal
J}^c_{n-1}+n\delta^{ab}
\delta_{n,1}(k+\alpha-\beta(k+5))\right]\Phi^b+
{\cal J}^b_{-1}\left[J^a_n,
\Phi^b\right].\ee
Now, for $n\geq0$,
\be\label{5.40}
\left[ J^a_n\Phi^b\right]|\Omega\rangle=\delta_{n0}
J_0^a\Phi^b|\Omega\rangle\ee
The r.h.s. of (\ref{5.40}) is evaluated by recalling
the explicit form of
$\Phi^b$ as given by (\ref{5.37}). We have from
(\ref{5.25}) and (\ref{5.5})
\bear\label{5.41}
\left(j^a_\psi\right)_0 g_j^{(0)}&=&-t_j^a g_j^{(0)}\nonumber\\
\tilde j^a_0(\tilde g_A)_{j-1}&=&-t^a_{j-1}(\tilde g_A)_{j-1}\ear
where $t^a_\ell$ are the $SU(2)$ generators in the spin-$\ell$
representation,
\be\label{5.42}
t^a_\ell=\frac{1}{(2\ell)!}\sum\left[\underbrace{t^a\otimes
\eins\otimes\ldots
\otimes\eins}_{2\ell}\right]_{\cal S}\ee
with $t^a\equiv t^a_{1/2}$, and $\eins$ the $2\times 2$
identity matrix; the
sum in (\ref{5.42}) runs over all possible permutations
of $t^a$ with the
identity matrices.

Using (\ref{5.41}) one has in the notation of (\ref{5.32}),
\be\label{5.43}
J_0^a\Phi^b=-tr\left( t^b t^a_j g^{(0)}_j\bar{\cal J}
(\tilde g_A)^{-1}_{j-1}\right)
+tr\left(t^b g^{(0)}_j\bar{\cal J}((\tilde g_A)^{-1}_{j-1}
t^a_{j-1})\right).\ee
One finds, after some algebra
\be\label{5.44}
tr\left(t^bt^a_jg^{(0)}_j\bar{\cal J}(\tilde g_A)
^{-1}_{j^{-1}}\right)=
f_{abc}\Phi^c+tr\left(t^bg_j^{(0)}\bar{\cal J}\left((\tilde
g_A)^{-1}_{j-1}t^a_{j-1}
\right)\right)\ee
Substituting (\ref{5.44}) into (\ref{5.43}) the contribution
of the second term in (\ref{5.43}) cancels,
leaving one with the expected
result
\be\label{5.45}
J^a_0\Phi^b=-f_{bac}\Phi^c\ee
Making use of this result we see that the operator (\ref{5.39}) will
annihilate the groundstate for all $n$, except $n=1$. For $n=1$ we
are left to compute
$f_{abc}J^c_0\Phi^b|\Omega\rangle$.

Proceeding as in (\ref{5.43}) and using (\ref{5.44}),
one evidently has
\be\label{5.46}
J^a_0\Phi^b=f_{abc}\Phi^c+(\beta-1)tr\left(t^bg_j^{(0)}
\bar{\cal J}((\tilde
g_A)^{-1}_{j-1}t^a_{j-1})\right)\ee
Contracting (\ref{5.46}) with $f_{abc}$, then replacing
$f_{abc}t^b$ in the
second term by $[t^c,t^a]$, and making use of
\be\label{5.47} \sum^3_{a=1}t^a_{ij}t^a_{k\ell}=-
\frac{1}{2}(\delta_{i\ell}
\delta_{jk}-\frac{1}{2}\delta_{ij}\delta_{k\ell})\ee
one obtains the remarkable result
\be\label{5.48}
f_{abc}{\cal J}^c_0\Phi^b=\left[-c_V+(\beta-1)(j-1)\right]\Phi^a\ee
Combining everything one thus finds
\be\label{5.49}
\left[Q,\Phi^{(k)}_{p,p-2}\right]|\Omega\rangle=
\left(-c_V+(\beta-1)(j-1)
+(k+\alpha-\beta(k+5))\right)
\eta_{-1}^a\Psi^a|\Omega\rangle\ee
BRST invariance thus requires
\be\label{5.50}
\alpha=\beta(j-k-6)+c_V+(j-1)-k\ee
This sample construction can be generalized to other primary
fields $\Phi_{p,q}^{(k)}$ along the lines discussed
following eq. (\ref{5.26}).

For $j=1$ $(p=3)$, expression (\ref{5.32}) can be written in the
compact form \cite{KZ} (recall (\ref{5.34})),
\be\label{5.51}
\Phi^{(k)}_{3,1}={\cal J}^a_{-1}\bar{\cal J}^{\bar
a}_{-1}tr\left(:(g^{(0)})^{-1}
t^ag^{(0)}t^{\bar a}:\right)\ee
where (see eq. (\ref{5.4}))
\be\label{5.52}
\left(g^{(0)}\right)^{ij}:=\left(g^{(0)}_{1/2}\right)^{ij}=
:e^{2\phi}
:\psi_2^{(0)i\alpha}\psi_2^{(0)\dagger j
\gamma}g_B^{\gamma\alpha}\ee
and use has been made of
\be\label{5.53}
g^{(0)}_{ij}\epsilon_{ik}=-\left(g^{(0)}\right)^{-1}_{kj}
\epsilon_{ji}\ee

\section{Conclusion}
In this paper we have shown how to obtain a fermionic coset realization
of the primaries in the minimal unitary models. In particular we have
been able to obtain in this framework an operator realization of all
the primaries in the Ising model, corresponding to the energy, order
and disorder operator, as well as the Onsager fermions.
The fermionic coset description played here a crucial role in the
realization
of the order-disorder algebra. The four-point functions of these
primaries were explicitly computed and shown to coincide with those
obtained from the representation theory of the Virasoro algebra
\cite{BPZ}.
This gave support to our identifications on operator level. In section 5
we then generalized the construction of the primaries $\Phi^{(1)}_{2,2},
\Phi^{(1)}_{2,1}$ of the
Ising model, to the primaries $\Phi^{(k)}_{p,q}$ with arbitrary
$k$ and $q=p,p-1$ and $p-2$. We have also indicated how the
construction would proceed for general $p-q>2$. In the general case
the evaluation of four-point functions will require the
knowledge of the four-point correlators of $SU(k)_{-2(k+1)}$ WZW fields.
In section 3 we have obtained a reduction formula reducing this
problem to the computation of the four-point function of $SU(2)_k$ WZW
fields, which are known \cite{FZ}.

We expect our general construction to prove useful for obtaining
a realization of the order-disorder algebra of other critical
statistical models. It may also prove useful in the study of statistical
models away from criticality.

\medskip
\noindent {\bf Acknowledgement:} One of the authors (D.C.C.) would
like to thank the Commission of the European Community for the
``Marie Curie Fellowship'', which made this collaboration possible.


\begin{thebibliography}{12}
\bibitem{BPZ} A. A. Belavin, A. M. Polyakov, and A. B. Zamolodchikov,
Nucl. Phys. {\bf B241},  (1981) 333.
\bibitem{Ibb} C. Itzykson, H. Saleur, and J.-B. Zuber,
Editors, ``Conformal Invariance and Applications to
Statistical Mechanics'', World Scientific 1988.
\bibitem{FQS} D. Friedan, Z. Qiu, and S. Shenker, Phys. Rev. Lett.
{\bf 52},  (1984) 1575.
\bibitem{H} D. A. Huse, Phys. Rev. {\bf B30}, (1984) 3908.
\bibitem{ABF} G. E. Andrews, R. J. Baxter, and P. J. Forrester, J. Stat.
Phys.
{\bf 35}, (1984) 193.
\bibitem{GKO} P. Goddard, A. Kent, and D. Olive, Int. J. Mod. Phys.
{\bf A1},  (1986) 303.
\bibitem{BRS} E. Bardacki, E. Rabinovici, and B. Saring, Nucl. Phys. {\bf
B229},
(1988) 151.
\bibitem{GK} K. Gawedzki and A. Kupiainen, Nucl. Phys. {\bf B320}, (1989)
624.
\bibitem{KS} D. Karabali, Q. Han Park, H. Schnitzer, and Z. Yang,
Phys. Lett. {\bf 216B},  (1989) 307;  D. Karabali and H. Schnitzer,
Nucl. Phys. {\bf B329}, (1990) 649.
\bibitem{CMR} D. Cabra, E. Moreno, and C. von Reichenbach, Int. J.
Mod. Phys. {\bf A5},  (1990) 2313.
\bibitem{KZ} V. G. Knizhnik and A. B. Zamolodchikov, Nucl. Phys. {\bf
B247}, (1984) 83.
\bibitem{FZ} A. B. Zamolodchikov and V. A. Fateev, Sov. J. Nucl. Phys.
{\bf 43}, (1986) 657.
\bibitem{NS} S. Naculich and H. Schnitzer, Nucl. Phys. {\bf B333}, (1990)
583;
{\bf B347},  (1990) 687.
\bibitem{IZ} C. Itzykson and J. M. Drouffe, {\it Statistical Field
Theory}, Vol. 1,2,
Cambridge Univ. Press 1989.
\bibitem{KC} L. P. Kadanoff and H. Ceva, Phys. Rev. {\bf B3},  (1971)
3918.
\bibitem{LP} A. Luther and I. Peschel, Phys. Rev. {\bf B12}, (1975) 3908.
\bibitem{CR} D. C. Cabra and K. D. Rothe, as ``Rapid Communication'' in
Phys. Rev. {\bf D51} (1995).
\bibitem{On} L. Onsager, Phys. Rev. {\bf 65}, (1944) 117.
\bibitem{P} A. M. Polyakov, in ``Fields, Strings and Critical Phenomena''
(Les Houches 1988, Eds. E. Brezin and J. Zinn-Justin, North Holland).
\bibitem{Wi} E. Witten, Comm. Math. Phys. {\bf 92},  (1984) 455.
\bibitem{PW} A. Polyakov and P. Wiegmann, Phys. Lett. {\bf 131B},  (1983)
121.
\bibitem{Fi} R. E. Gamboa Saravi, F. A. Schaposnik, and J. E. Solomin,
Nucl. Phys. {\bf B185},  (1981) 238.
\bibitem{BS} B. Schroer and T. T. Truong, Nucl. Phys. {\bf B154}, (1979)
125.
\bibitem{DFSZ} P. Di Francesco, H. Saleur, and D. B. Zuber, Nucl. Phys.
{\bf B29} [FS20],   (1987) 527.
\bibitem{He} M. Henneaux, Phys. Rep. {\bf 126}, (1985) 1.
\bibitem{Ta} Y. Tanii, Mod. Phys. Lett. {\bf A5}, (1990) 927.
\bibitem{Ba} F. Bastianelli, Nucl. Phys. {\bf B361}, (1991) 555.
\end{thebibliography}
\end{document}